\documentclass[lettersize,journal]{IEEEtran}
\usepackage{amsmath,amsfonts}
\usepackage{algorithmic}
\usepackage{algorithm}
\usepackage{array}
\usepackage[caption=false,font=normalsize,labelfont=sf,textfont=sf]{subfig}
\usepackage{textcomp}
\usepackage{stfloats}
\usepackage{url}
\usepackage{verbatim}
\usepackage{graphicx}
\usepackage{cite}
\usepackage{hyperref}
\usepackage{mathtools}
\usepackage{booktabs}
\usepackage{threeparttable}
\usepackage[table]{xcolor} 
\usepackage{soul, color}
\usepackage{todonotes}

\soulregister\cite7
\soulregister\ref7
\soulregister\pageref7

\hypersetup{
  breaklinks   = true, 
  colorlinks   = true,    
  urlcolor     = blue,    
  linkcolor    = blue,    
  citecolor    = gray      
}

\begin{document}

\title{Exploring the Relationship Between Ownership and Contribution Alignment and Code Technical Debt}

\author{Ehsan Zabardast, 
        Javier Gonzalez-Huerta,
        Francis Palma,
        Panagiota Chatzipetrou
\thanks{E. Zabardast and J. Gonzalez-Huerta are with Software Engineering Research Lab (SERL), Blekinge Institute of Technology, Sweden.}
\thanks{F. Palma is with University of New Brunswick, Canada.}
\thanks{P. Chatzipetrou is with University of \"Orebro, Sweden.}
}

\markboth{Journal of IEEE\ Class Files,~Vol.~XX, No.~X, April~2023}%
{Zabardast \MakeLowercase{\textit{et al.}}: A Sample Article Using IEEEtran.cls for IEEE Journals}


\maketitle

\begin{abstract}
\textbf{Context:} Software development organisations strive to remain effective and efficient while the complexity of the systems they develop continues to grow. To tackle this challenge, they are often organised into small teams working on separated components that can be independently developed, tested, and deployed. Effective alignment between architecture and organisational structures is essential for effective communication and collaboration and helps teams reduce code and architectural degradation. Assigning responsibility for specific components to the teams that primarily work on them is crucial to achieving these goals. 

{\color{blue}
\textbf{Objective:} This article reports a study aiming to explore the relationship between ownership and contribution alignment (contribution alignment, for short) and code quality degradation, measured as technical debt per line of code. It also examines how changes in team structure impact their ability to manage code quality degradation.
}

\textbf{Method:} We have conducted a case study in a company developing a large software system, analysing ten components managed by one team. This team was later split into two, redistributing their components between the two new teams. We have collected archival data from the development tools used in their daily operations.


\textbf{Results:} {\color{blue}Before the split, there was a statistically significant negative correlation between contribution alignment and technical debt density in four components, indicating that higher contribution alignment leads to lower technical debt density and better code quality. After the split, this negative correlation persisted in three components, while five components showed a positive correlation, suggesting that low contribution alignment might exacerbate code quality degradation.}

\textbf{Conclusion:} Our findings suggest that contribution alignment can be important in controlling code quality degradation in software development organisations. 
{\color{blue}Ensuring teams are responsible for components they are most familiar with and minimising dependencies between teams can help mitigate code quality degradation.}

\end{abstract}


\section{Introduction}\label{sec1:introduction}

Large-scale software organisations tend to organise their systems as a constellation of components that are usually developed and maintained by different teams as a way of ``componentising'' their software architectures. Microservices architecture style~\cite{Jamshidi2018,FowlerMicroservices2014} is a widely adopted specific example. In a microservices architecture, applications are composed of many small, independent services that communicate with each other, and that are owned, developed, and maintained independently by different teams\cite{NewmanMonolithToMicro2019}. This approach can help organisations to enhance maintainability, improve agility, and reduce time-to-market \cite{Newman2015}. At the same time, its adoption introduces challenges, such as the need for communication and coordination between development teams, and to manage dependencies among microservices and the teams developing them \cite{NewmanMonolithToMicro2019,Newman2015}. 

{\color{blue}
Although there are several approaches to handling ownership, with different effects on teams' and individuals' autonomy, large-scale organisations usually rely on the \textit{weak ownership principle}~\cite{FowlerOwnership2006}, which consists of one team \textit{owning} a component (or a microservice). The \textit{owning} team is responsible for the quality of the \textit{owned} component, and its members usually decide about code changes made by members of other teams through code reviews~\cite{FowlerOwnership2006}. However, this responsibility does not necessarily mean they are the sole contributors to the code of that component, not even the \textit{main} contributors.}

Prior studies explored the relationship between the number of developers in open-source projects and their quality~\cite{Meneely2009,Schweik2008}.  However, having multiple authors who belong to multiple clusters might have a negative impact on quality, for example, the introduction of security flaws~\cite{Meneely2009}. The quality decline associated with authors from different clusters can be attributed to the concept of \textit{responsibility diffusion}. Responsibility diffusion~\cite{Barley1968} is a concept from social sciences that describes the lack of action when many people witness a criminal action. All witnesses tend to think someone else will take action (i.e., call the police), and yet, no one does. In software engineering, we associate the concept of responsibility diffusion with the lack of sense of responsibility when a given component (or code element) has multiple authors belonging to different organisations or clusters, and therefore no one will take corrective actions~\cite{Tornhill2015}. This article goes one step forward, claiming that this lack of sense of responsibility is also affected by the fact that an owning team does not have a major contribution role to the owned code element or component, i.e., the ownership and contribution alignment degree is low.

In proprietary software systems, the alignment between the architecture and the organisation's communication structure becomes a critical factor to success in the development of large, very large, and ultra-large-scale\footnote{Based on the definitions by~\cite{Digngsor2014} and~\cite{Northrop2006}, respectively.} software systems \cite{baskarada2020,Newman2015}. Conway \cite{conway1968committees} hypothesises that \textit{``organisations that design systems tend to produce designs that mirror the communication structure of these organisations.''} It seems natural that when we componentise the architecture, the team constellation and the ownership should be adapted to minimise communication overhead and the dependencies between teams at the task level (i.e., a team depending on some other team's work to finish a task). {\color{blue} Ideally, a team will be more productive if it is responsible for the components to which they are the main contributors. This aligns the team's responsibilities with their contribution, which might help reduce the average time to review and integrate new code~\cite{thongtanunam2016revisiting}, and help to mitigate the code quality degradation.} 

{\color{blue}
In this paper, we use Technical Debt Density (TDD), defined as the total amount of Technical Debt reported by SonarQube normalised per line of code, as a proxy for code quality. Although TD is not always synonymous with code quality, many widely-adopted static code analysis tools report quality problems using the TD construct~\cite{Avgeriou2016}. Technical Debt refers to the long-term costs of maintaining and updating a software system due to suboptimal or inefficient solutions used during development~\cite{avgeriou2023technical}. Tools like SonarQube report the TD as the amount of time needed to remediate code quality issues. TDD is a metric that has been used in other studies on the field \cite{zabardast2022impact,digkas2020can,al2019evolution}.}





{\color{blue}We are conducting an observational study; therefore, there are no interventions. Our goal is to identify meaningful patterns~\cite{ayala2021use}. We hypothesise that a low contribution alignment degree can trigger faster degradation of code quality.} Teams with low contributions to a given component might not perceive it as their responsibility and might not have enough knowledge to take corrective actions or decide on changes implemented by others.

To study this phenomenon, i.e., the impact that the degree of alignment between ownership and contribution might have on code quality degradation, we have conducted a case study on a software company developing a very large software system (1.5 million LOC, developed by $>$20 teams). We have followed the evolution of ten software components (micro-services) initially developed by one team over three years. During this period, the team was split into two, and the components owned by the team were distributed between the new teams. The degree of contribution alignment after the split changed substantially for some components since some of the main contributors were not in the owning team anymore. {\color{blue}Therefore, an additional goal of this case study is to investigate how sudden changes in team constellation might affect teams' effectiveness in addressing code quality degradation in the components they are responsible for.}

The remainder of the article is structured as follows: Section~\ref{sec6:related_work} discussed the related work. Section~\ref{sec2:research_methodology} describes the research methodology, by describing the case, the data collection, and the analysis methods. Section~\ref{sec3:results} presents the results. Section~\ref{sec4:discussion} discusses the main findings and implications. Section~\ref{sec5:threats_to_validity} discusses and addresses the limitations and threats to validity. Finally, Section~\ref{sec7:conclusion} presents the conclusion and future work.

\section{Related Work}\label{sec6:related_work}

There are several research works that scrutinise the relationship between team or organisation structure and software quality \cite{thongtanunam2016revisiting,posnett2013dual,bird2011don,rahman2011ownership,Meneely2009,nagappan2008influence,Schweik2008,brooks1974mythical}, some of which directly put the focus on the relationship of ownership and software quality, e.g., \cite{thongtanunam2016revisiting,posnett2013dual,bird2011don,rahman2011ownership}. The concept of TDD \cite{leven2022broken,arvanitou2022practitioners,benidris2021technical,digkas2020can,al2019evolution} has been used as a construct to reason about the evolution of code and architectural degradation~\cite{Zabardast2022Assets}. 
The degree to which developers contribute to a particular project has also been addressed in several studies, e.g.,  \cite{oliveira2020code,de2018measuring,parizi2018measuring,gousios2008measuring} as a way to characterise the degree in which developers participate in software projects. 

In the following subsections, we summarise relevant works in each of these areas to position the contribution of our study in relation to the existing literature on the field.

\subsection{Organisations, Team Structure, and Quality}

The quality of any product is strongly affected by organisation structure \cite{brooks1974mythical}. This is empirically shown by Nagappan et al. \cite{nagappan2008influence} on the relationship between organisational structure and software quality via eight organisational complexity metrics from the code viewpoint. Examples include ``\textit{the absolute number of unique engineers who have touched a binary and are still employed by the company}'' and ``\textit{the total number of unique engineers who have touched a binary and have left the company as of the release date of the software system}'' \cite{nagappan2008influence}. The authors develop a model to predict the failure-proneness and achieve a precision of up to 86.2\% and recall of up to 84\%, compared to other traditional metrics, including code churn and code complexity. Thus, organisational and owner metrics are more effective in software quality estimation.

There also have been studies analysing the impact that the number of developers might have on the quality of Open-Source Systems (OSS) (e.g., \cite{Meneely2009,Schweik2008}). Schweik et al.~\cite{Schweik2008} opposes two laws (or principles): Brooks' law~\cite{brooks1974mythical} against Linus' law~\cite{Raymond1999}. On the one hand, Brooks' law states that adding more (human) resources to a software project tends to make it late. On the other hand, Linus' law~\cite{Raymond1999} named after the creator of the Linux operating system, states that ``with enough eyeballs, all bugs are shallow''. In other words, adding more developers (in OSS) encompasses higher quality. In \cite{Schweik2008}, authors report the impact of the number of developers on the ``survivability'' of OSS projects, using the latter as a proxy of its quality. Meneely and Williams~\cite{Meneely2009} did a similar study, in this case just focusing on Linus' law, specifically regarding the number of vulnerabilities. The main result is that the presence of developers from different clusters seems to have a negative impact on quality in the form of more vulnerabilities present in the system.

\subsection{Software Ownership and Quality}

Bird et al., \cite{bird2011don} examine the relationship between various (software) ownership measures (e.g., \textit{the number of low-expertise developers} and \textit{proportion of ownership for the top owner} - although in this case, the authors address individual contributors and not teams) and software failures, and find that the measures of ownership are associated with pre-release faults and post-release failures. In particular, the evidence of correlation is significant for minor contributors, i.e., those who contribute less and infrequently on a module. Our \textit{contribution degree} calculation uses the degree of contribution of the owning team, instead of the top contributor, as the metric to reason about the degree of alignment between ownership and contribution, and its impact on TDD.

Thongtanunam et al. \cite{thongtanunam2016revisiting} argue that code ownership comes with responsibility, and developers who author the majority of changes to a module are presumably the owners. However, contribution may come in other forms, e.g., in modern code review by criticising code changes authored by other developers. In our \textit{contribution degree} calculation, we consider code review information. The authors also show that  67\%-86\% of the developers who contribute to a module do not author code changes, i.e., contribute by reviewing code. Among them, only 18\%-50\% are core team members. Thus, the authors suggest that code review should be included in code ownership estimation. While evaluating the relationship between review-specific and review-aware code ownership and defect-proneness, the authors find that developers with low traditional and review ownership are more prone to post-release faults.  These results aligned with our findings that suggest that higher ownership might help teams keep TDD\footnote{One of the dimensions used by SonarQube, which includes bugs, as detected using static analysis, which is different from the number of bug reports in ticket management systems.}.


Moreover, Posnett et al.~\cite{posnett2013dual} argue that low ownership of a module (i.e., too many contributors) can affect code quality. The authors define a metric called DAF (Developer’s Attention Focus, which measures the focus of a developer on some activities) and show that more focused developers introduce fewer defects than less-focused developers. In contrast, files receiving narrowly focused activity are more likely to have faults than others.

{\color{blue}In a recent study, Sundelin et al.~\cite{Sundelin2024} analyse the impact that ownership might have in the introduction of a very specific technical debt item, i.e., code clones. The authors built a Multi-Level Generalised Linear Bayesian Model to estimate the number of introduced code clones for a given change by a given team. The results show that the team, and whether they own a given component, plays a role in how code clones are introduced, but it is still unclear whether the misalignment between ownership and contribution actually has an impact on the introduction of technical debt and with a broader perspective. Although the authors use the original OCAM Ownership Model~\cite{zabardast2022} they only use it to illustrate the degree to which a given team has contributed to the repositories, but they do not consider the temporal evolution of ownership over time nor they try to establish a relationship between ownership alignment and technical debt.}



\subsection{Technical Debt Density in Source Code}

In an exploratory study, Al Mamun et al. \cite{al2019evolution} investigates the ability of the `TDD trend' metric to estimate the evolution of technical debt (TD). The `TDD trend' is the slope of two consecutive `TDD' measures. Using the TDD trend' metric, the authors observe that a file has the highest level of TDD at the initial stage of its revisions, and TDD decreases as the file size increases.

In another study, Digkas et al.\cite{digkas2020can} examine the relationship between the amount of technical debt in new code and the evolution of technical debt in a system. The authors argue that TD grows in absolute numbers as software systems evolve. In contrast, the density of TD (i.e., TD divided by lines of code, also known as normalised TD) may decline due to refactoring or the development of new artefacts with less TD. As their findings, the authors report that among the three major types of code changes (insertion, deletion, and modification), the contribution of code modification (i.e., refactoring) is strongly related to the change in the TDD.


Lev\'en et al.~\cite{leven2022broken} investigate the causal relationship between the existing TDD of a system and developers' tendency to introduce new TD during the system's evolution (using the Broken-Window Theory). The findings suggest that existing TD affects developers' tendency to introduce new TD during further system development.

Arvanitou et al.~\cite{arvanitou2022practitioners} investigate how the accumulation of TD can be explained in (scientific) software development. To do that, authors first identify software engineering practices used to develop scientific software and the most common causes of introducing TD and then map them. Findings suggest that the scientists that develop scientific software lack certain software engineering practices and introduce TD in scientific software. To minimise the TD in scientific software, the authors recommend reusing software libraries and process improvement methodologies and working in pairs (i.e., applying eXtreme Programming), which would minimise about half of the causes of introducing TD.

\subsection{Developer Contribution Metrics and Quality}

Developers' contributions can be estimated and likewise can be linked to the quality of the source code. De Bassi et al. \cite{de2018measuring} argue that software quality can be directly linked to the volume of collaboration and commitment in the development team. The authors evaluate 20 quality metrics related to complexity, inheritance, and size that can measure team members’ participation regarding their source code contribution to the project. In the end, those metrics are analysed based on positive, negative, and no influence on source code quality, including maintainability, testability, and understandability.

Parizi et al. \cite{parizi2018measuring}, utilising git-driven technology and its features, estimate and visualise a team member’s contribution using a set of metrics, including \textit{the number of commits}, \textit{number of merge pull requests}, \textit{number of files}, and \textit{total lines of code}. For computing these metrics, authors primarily rely on git logs as inputs to extract the performance data in combination with the total time spent on a project each day.

In another work, Oliveira et al. \cite{oliveira2020code} define source code ownership by a developer when they contributed most to a source code file. The authors study several code-based and commit-based developer productivity metrics, including \textit{source lines of code by time} (SLOC/Time, code-based) -- the larger the source code created by the developer, the higher their productivity; and \textit{commits performed by time} (Commits/Time, commit-based) -- the higher the number of commits, the higher is their productivity. Using those metrics, the authors explore whether and to what extent developer productivity metrics have a relationship with developers' productivity. Authors find that code-based metrics better explain productivity than commit-based metrics.

The literature shows that in software development, team structure, and the developers' contribution seem to impact quality. However, what is not known is whether the misalignment between ownership and contribution (i.e., low values of \textit{contribution degree}) affects TD. This article aims to bridge this gap by studying the relationship between the TD and the misalignment between ownership and contribution. To measure the misalignment between ownership and contribution, we include several perspectives, including contributions to code reviews as suggested by~\cite{thongtanunam2016revisiting}, and measure contribution degree based on the degree of contribution by the owning team, as suggested by~\cite{bird2011don}.

\section{Research Methodology}\label{sec2:research_methodology}

{\color{blue}
In this article, we are addressing the following research question: \textbf{RQ:} \textit{How does the degree of ownership and contribution alignment impact the degradation of code quality?}
}

\begin{figure}[ht]
\centering
  \includegraphics[width=.5\textwidth]{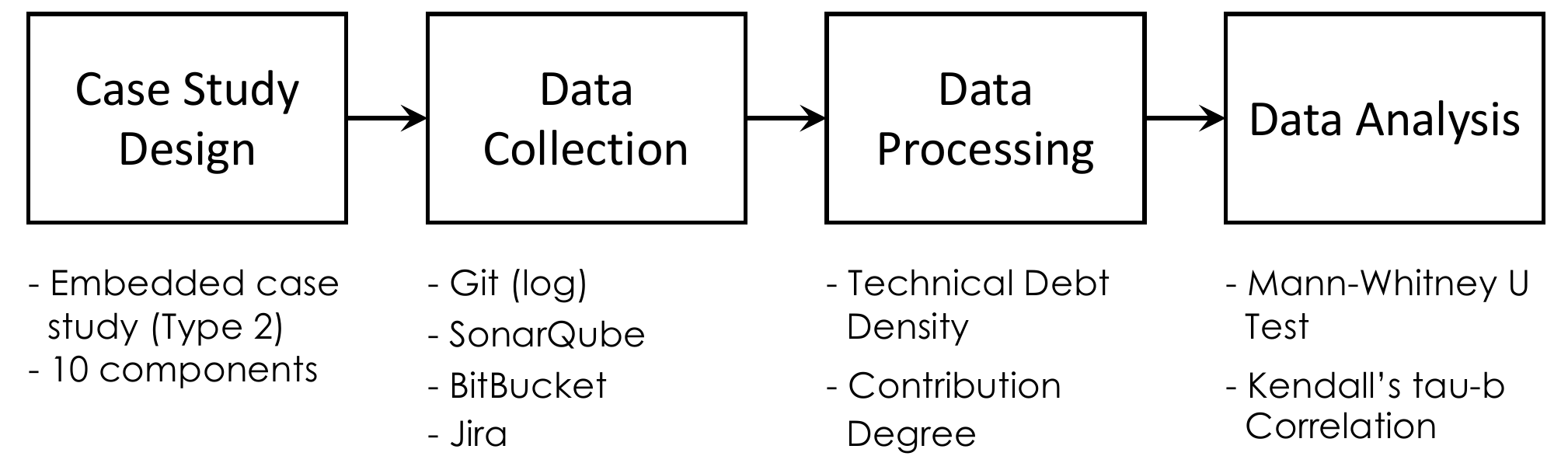}
\caption{The research method.}
\label{fig:rm_steps}      
\end{figure}

To address the research question, we have designed an embedded case study (type 2) according to the definition by Yin \cite{yin2018case}. 
We conduct this study in an industrial setting and rely on the analysis of ten \texttt{Java} components developed by a company that has chosen to remain anonymous.  We use archival data collected from the tools that the company uses during development. The research approach of this article is illustrated in Figure \ref{fig:rm_steps}.

\subsection{Case Study Design}

{\color{blue}
The \textit{context} of the case study is a large-scale software development company that develops cloud-based financial and accountancy services. It is a mature company in its development practices and has well-established, successful products on its portfolio, and it has been selected by convenience (availability and access), following the information-oriented principle~\cite{Flyvbjerg2006}. The company is interested in continuously improving its products and ways of working. Therefore, it was willing to participate in the study and learn from its results. The \textit{case} is a software system developed by the company with 1.5 million LOC developed by $>$20 teams. \textit{Units of analysis} are the ten components, i.e., microservices, that are part of this software system. Table \ref{tab:component_info} presents the details of the investigated components and their owning teams in this study.
}
\begin{table*}[ht!]
\caption{Component information - size, number of commits, average cyclomatic complexity of the component (2020-2023), number of active development weeks, and the owning team including members' seniority (average time in the company).}
\label{tab:component_info}
\centering
\rowcolors{2}{gray!25}{white}
\begin{tabular}{cccccc}
\textbf{Component} & \textbf{Size (LOC)} & \textbf{Commits} & \textbf{Avg. cyclomatic complexity} & \textbf{Active weeks} & \textbf{Team (avg. seniority - years)} \\
\hline
C1 & 18,244& 1,019 & 1307 & 122 & Brown (4.1) \\
C2 & 13,968 & 714 & 1034 & 89 & Gray (2.9)\\
C3 & 12,290 & 872 & 866 & 105 & Gray (2.9)\\
C4 & 5,019 & 192 & 512 & 50 & Brown (4.1) \\
C5 & 11,001 & 691 & 761 & 70 & Gray (2.9)\\
C6 & 7,642 & 366 & 526 & 74 & Brown (4.1) \\
C7 & 31,708 & 1,136 & 2047 & 130 & Brown (4.1) \\
C8 & 1,872 & 203 & 140 & 57 & Gray (2.9)\\
C9 & 11,994 & 434 & 868 & 78 & Brown (4.1) \\
C10 & 17,187 & 617 & 1084 & 101 & Brown (4.1) \\
\hline
\textbf{Total:} & \textbf{130,925} & \textbf{6244} & - & - & - \\
\hline
\end{tabular}
\end{table*}

The analysis period is three years from January 2020 to December 2022. The measurements are collected in weekly intervals, and we only consider data from the weeks where there was active development on the component. Choosing shorter intervals, e.g., days, or even individual commits, results in a high proportion of observations with \texttt{zero} values which might impact the variables under study.

\begin{figure}[ht]
\centering
  \includegraphics[width=.45\textwidth]{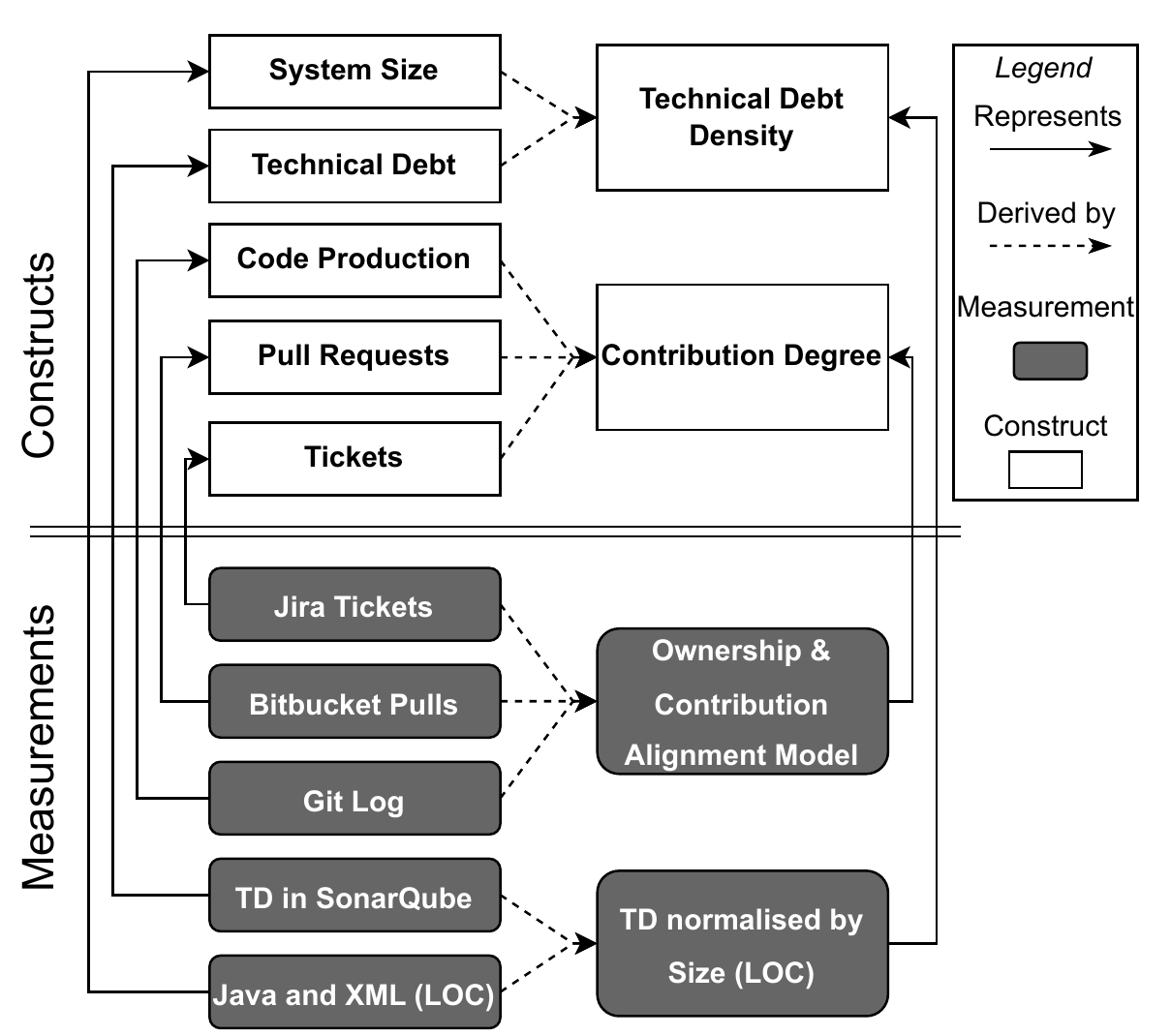}
\caption{Study constructs and measurements.}
\label{fig:construct_measurement}      
\end{figure}

In the following, we describe the main constructs and measurements used for the quantitative analysis of the data collected in this study. We use \textit{Degree of Ownership and Contribution Alignment}, and \textit{Technical Debt Density} (TDD \cite{digkas2020can,al2019evolution}, {\color{blue} as a measure for code quality,}  as the main constructs. Figure~\ref{fig:construct_measurement} illustrates the constructs used in this work.

{\color{blue}
The two main variables used in this article are: \textit{Contribution Degree} and \textit{Technical Debt Density (TDD)}. Their definitions are:

\begin{itemize}
    \item The \textit{Degree of Ownership and Contribution Alignment} or \textit{Contribution Degree} represents how much of the contribution to a component comes from the officially assigned \textit{owner}. A 100\% contribution degree means that all the contribution to a component is from the formally responsible team (owning team), as designated by the organisation. From this point on, we will refer to the \textit{degree of ownership and contribution alignment} as \textit{Contribution Degree}. We use a modified ownership and contribution alignment model (OCAM)~\cite{zabardast2022} to calculate the \textit{Contribution Degree}. Instead of using the \emph{ranks} provided by the original model, we use \emph{ratios}. The contribution degree is the ratio of the teams' contribution for each measure and is averaged between all measures for the final result, and subsequently, used in this study (as presented in the Figures \ref{fig:box_contribution} and \ref{fig:result_tdd_evo}). In this article, we are using an extension of the original OCAM model reported in~\cite{zabardast2022}. We use four measures out of the seven presented in the original model, but we compile them in a single metric with a ratio scale that allows us to use it as contribution degree, i.e., the ratio of contribution to a component by the owning team. The contribution degree is calculated using the contribution to \texttt{code production}, \texttt{pull requests}, and \texttt{tickets}. `Code production' is the degree of contribution of a team both in terms of the number of commits and code churn \cite{munson1998code}. `Pull requests' is the degree of contribution of a team in terms of the number of created pull requests \cite{thongtanunam2016revisiting}. `Tickets' is the degree of contribution of a team in terms of the number of tickets created on the ticket management system. The results obtained from applying the OCAM model to the components were validated with component owners and their managers before conducting the statistical tests. 
    \item Technical Debt Density (TDD) is the normalised remediation time, i.e., technical debt as reported by SonarQube, per line of code~\cite{digkas2020can,al2019evolution}. TDD is calculated by dividing the total amount of TD by the component size (LOC).
\end{itemize}

$ContributionDegree(t)$ (See Equation \ref{ocam_equation}) and $TDD(t)$ (See Equation \ref{tdd_equation}) are both defined with respect to a time period: 

}

\begin{equation}
    \label{ocam_equation}
    Contribution.Degree(t)=\frac{C(t)+Ch(t)+P(t)+T(t)}{4}
\end{equation}

\noindent With $C(t)$ being the contribution to the commits (See Equation~\ref{eq_c}), i.e., the percentage of commits done by the team on a specific component; $Ch(t)$ being the contribution to code churn (See Equation \ref{eq_ch}), i.e., the percentages of code written by the team in the code base of a specific component; $P(t)$ being the contribution to pull requests (See Equation \ref{eq_p}), i.e., the percentage of pull requests created by the team on a specific component; and $T(t)$ being the contribution to tickets (See Equation \ref{eq_t}), i.e., the percentage of tickets created by the team on a specific component.

\begin{equation}
    \label{eq_c}
    C(t)=Commit_{Cont.Deg.}(t)=\frac{Commits_{team}}{Commits_{all}}\times100
\end{equation}

\begin{equation}
    \label{eq_ch}
    Ch(t)=CodeChurn_{Cont.Deg.}(t)=\frac{CodeChurn_{team}}{CodeChurn_{all}}\times100
\end{equation}

\begin{equation}
    \label{eq_p}
    P(t)=PullRequest_{Cont.Deg.}(t)=\frac{PullRequests_{team}}{PullRequests_{all}}\times100
\end{equation}

\begin{equation}
    \label{eq_t}
    T(t)=Ticket_{Cont.Deg.}(t)=\frac{Tickets_{team}}{Tickets_{all}}\times100
\end{equation}


\noindent $TDD(t)$ is defined as the amount of TD $t$ normalised by the component size at a given point in time. It is operationalised as the total remediation time for all TD items present in the component on a given instant $t$, as reported by SonarQube, divided by the component size at $t$: 

\begin{equation}
    \label{tdd_equation}
    TDD(t)=\frac{\sum_{i=1}^{\#SystemTDI(t)}(RemeditionTime_i)}{Size(t)}
\end{equation}

\subsection{Data Collection and Processing}\label{sec2:data_collection}

We used widely used tools such as \texttt{Git} and API endpoints to access the tools used by the company in their daily operations. We collected data from \texttt{Git}\footnote{https://git-scm.com}, \texttt{BitBucket}\footnote{https://bitbucket.org}, \texttt{Jira}\footnote{https://www.atlassian.com/software/jira}, and \texttt{SonarQube}\footnote{https://www.sonarsource.com/products/sonarqube/}. We collected the developer team affiliation data from the company's API endpoint. Here, we present how the data was collected from each tool. 

\begin{itemize}

    \item \texttt{Git:} We used \texttt{Git Log} to collect the data regarding the component size and code production to calculate commit frequency and code churn. Component size is calculated using CLOC\footnote{https://cloc.sourceforge.net}. The historical data was collected by checking out the commits and calculating the size using CLOC. When calculating the size of the system, we include only the \texttt{Java} and \texttt{XML} files. We exclude the other files since technical debt is calculated based on \texttt{Java} and \texttt{XML} profiles.

    \item \texttt{BitBucket:} We used the BitBucket API to collect the data regarding the creation of pull requests. The raw data was collected using the company API endpoint.

    \item \texttt{Jira:} We used the Jira API to collect the data regarding the creation of tickets. The raw data was collected using the company API endpoint.
    
    \item \texttt{SonarQube:} We used SonarQube to calculate the amount of code technical debt for each component. The data gathered from SonarQube are the issues detected by the quality profile and quality gate configured by the company. The data is collected via the company's SonarQube instance API endpoint. SonarQube has been used in similar studies on the topic of TD, e.g., \cite{siavvas2022technical,zabardast2022impact,digkas2020can,lenarduzzi2020sonarqube,zabardast2020,lenarduzzi2019diffuseness,digkas2018developers,digkas2017evolution}. The tool provides the estimated remediation time (i.e., effort in minutes) required to resolve TD items (issues in the SonarQube terminology). The accumulated TD of a component at a point in time is the total remediation time for all the issues detected in its codebase. We considered TD as \textit{repaid} when issues are tagged as `fixed' or `closed' or are removed from the component\footnote{The company's SonarQube instance has been configured not to remove technical debt items records when they are removed without tagging them as fixed or closed, but to keep their introduction and removal dates to improve the reliability of the data.}.
    
\end{itemize}

\subsection{Data Analysis}\label{sec2:data_analysis}

We used the Shapiro-Wilk test as the normality test to analyse whether \textit{contribution degree} and TDD `before' and `after' the team split were normally distributed\footnote{All the statistical tests reported in this subsection were conducted using \texttt{Python ver. 3.9.9} and \texttt{NumPy ver. 1.10.1}}. We decided to proceed with non-parametric tests as the data was normally distributed only in one case.

\subsubsection{Kendall's tau-b Correlation Test}

To test whether there is an association between \textit{Contribution Degree} and TDD and the type of association, we used Kendall's tau-b correlation coefficient\footnote{The correlation tests were conducted using \texttt{IBM SPSS Statistics ver. 28}}. Kendall's tau-b correlation coefficient is a non-parametric measure of the strength and direction of association between \textit{Contribution Degree} and TDD. Kendall's tau-b correlation coefficient can range from one (1) to minus one (-1). The closer a result is to one, the higher the level of association. The corresponding p-values for Kendall's tau-b correlation coefficients indicate the presence of a significant relationship between \textit{Contribution Degree} and TDD. Kendall's tau-b correlation coefficient is more robust and efficient than alternative non-parametric tests (i.e., Spearman rank correlation), and it is preferred when we have smaller samples. 

\vspace{1mm}
\noindent\textbf{Variables}: Contribution degree for a give component \textit{in that particular week} and TDD for a given component \textit{in that particular week}.

\begin{quote}
    \textit{Hypothesis} $H_{0}$: There is no association between contribution degree and TDD.
    
    \textit{Hypothesis} $H_{1}$: There is an association between contribution degree and TDD. 
\end{quote}

\subsubsection{Mann-Whitney U Test}

{\color{blue} To test if splitting the team impacts the \textit{contribution degree}, we used the Mann-Whitney U test \cite{mann1947test,lehmann1975nonparametrics} to determine whether the \textit{contribution degree} is different before and after the team split. In other words, we wanted to test whether the team split is a confounding factor that impacts the \textit{contribution degree}. If there is a statistically significant difference before and after the team split,} the analysis of the relationship between \textit{contribution degree} and TDD needs to be done separately.

\vspace{1mm}
\noindent\textbf{Variables}: The independent variable is the categorical groups of `before' and `after' the split for a given component. The dependent variable is the Contribution Degree for a given component \textit{in that particular week}.

\begin{quote}
    \textit{Hypothesis} $H_{0}$: There is no difference in contribution degree between `before' and `after' teams split.
    
    \textit{Hypothesis} $H_{1}$: There is a difference in contribution degree between `before' and `after' teams split.
\end{quote}

To test if the change in team structure impacts the accumulation of TDD, we used the Mann-Whitney U test. We used the test to examine whether the TDD is different between `before' and `after' the team split. 
    
\vspace{1mm}
\noindent\textbf{Variables}: The independent variable is the categorical groups of `before' and `after' the split for a give component. The dependent variable is TDD for a given component \textit{in that particular week}.

\begin{quote}
    \textit{Hypothesis} $H_{0}$: There is no difference of TDD between `before' and `after' teams split.
    
    \textit{Hypothesis} $H_{1}$: There is a difference of TDD between `before' and `after' teams split.
\end{quote}

\subsection{Validation of the Results}

During the execution of the case study, the research team meets the teams under study in recurring meetings every two weeks to report the plan and discuss this and other studies, as well as in focus groups in which we presented the results and collected their feedback. The results of the \textit{contribution degree} were presented to the development team, the engineer leader and product owners to validate the accuracy of our calculations, and whether they reflected the events that occurred during the studied period (e.g., top contributors leaving the organisation). 

Moreover, we gathered data regarding potential explanations for some sudden changes in TDD during the studied period. In this latter case, to avoid introducing bias, we did not reveal the purpose of the analysis until presenting the full results. This way, we avoid the studied teams trying to defend suboptimal decisions. For this data gathering sessions, we sometimes visualised both non-normalised TD and TDD to reason about its evolution.

\section{Results}\label{sec3:results}

This section presents the results investigating and exploring the impact of ownership and contribution alignment on code technical debt for ten components. Table \ref{tab:descriptive_stats} presents the descriptive statistics for \textit{Contribution Degree} and TDD.

\begin{table*}[ht!]
\caption{Descriptive statistics for contribution degree and TDD. The table is split to present the data separately for before and after the team structure change. $N$ is the total number of weeks before and after the split.}
\label{tab:descriptive_stats} 
\rowcolors{2}{gray!25}{white}
\centering
\begin{tabular}{lccccc|ccccc|c}
\rowcolor{white}\multicolumn{12}{c}{\textbf{Contribution Degree}}\\
\hline
 & \multicolumn{5}{c|}{\textbf{Before}} & \multicolumn{5}{c|}{\textbf{After}}&\textbf{Total N}\\
 \hline
\rowcolor{white}& N & Mean & STD & Min & Max & N & Mean & STD & Min & Max & \\
\hline
C1  & 46 & 53.183 & 9.372 & 31.3 & 58.9 & 76 & 53.271 & 12.969 & 36.4 & 75.9 & 122 \\
C2  & 32 & 68.516 & 10.502 & 52.1 & 82.3 & 57 & 41.853 & 8.535 & 32.8 & 82.4 & 89 \\
C3  & 38 & 84.482 & 9.462 & 68.2 & 97.2 & 67 & 35.278 & 18.069 & 20.8 & 81.7 & 105\\
C4  & 10 & 38.240 & 2.370 & 36.0 & 42.5 & 40 & 19.230 & 12.550 & 2.8 & 32.8 & 50\\
C5\textdagger  & -  & -      & -     & -    & -    & 70 & 83.651 & 3.672 & 73.9 & 89.3 & 70 \\
C6  & 22 & 35.259 & 3.879 & 22.6 & 39.4 & 52 & 37.596 & 19.230 & 11.1 & 55.8 & 74\\
C7  & 45 & 42.258 & 5.502 & 29.2 & 45.3 & 85 & 40.935 & 15.336 & 18.0 & 62.9  & 130\\
C8\textdagger  & 3  & -      & -     & -    & -    & 54 & 31.924 & 32.504 & 7.3 & 81.4 & 57 \\
C9  & 26 & 58.127 & 7.181 & 42.4 & 65.7 & 52 & 51.398 & 7.071 & 39.5 & 59.7 & 78\\
C10 & 36 & 64.533 & 2.632 & 57.9 & 66.9 & 65 & 47.646 & 8.094 & 28.8 & 55.8 & 101\\
\hline
\rowcolor{white}\multicolumn{12}{c}{\textbf{}}\\
\rowcolor{white}\multicolumn{12}{c}{\textbf{Technical Debt Density}}\\
\hline
 \rowcolor{white}& \multicolumn{5}{c|}{\textbf{Before}} & \multicolumn{5}{c|}{\textbf{After}} & \textbf{Total N}\\
 \hline
\rowcolor{white}& N & Mean & STD & Min & Max & N & Mean & STD & Min & Max & \\
\hline
C1  & 46 & 0.221 & 0.006 & 0.211 & 0.231 & 76 & 0.205 & 0.004 & 0.193 & 0.213 & 122 \\
C2  & 32 & 0.232 & 0.051 & 0.167 & 0.392 & 57 & 0.158 & 0.028 & 0.117 & 0.203 & 89\\
C3  & 38 & 0.093 & 0.011 & 0.080 & 0.126 & 67 & 0.099 & 0.014 & 0.079 & 0.112 & 105\\
C4  & 10 & 0.085 & 0.004 & 0.079 & 0.089 & 40 & 0.085 & 0.008 & 0.072 & 0.094 & 50\\
C5\textdagger  & -  & -     & -     & -     & -     & 70 & 0.025 & 0.036 & 0.001 & 0.101 & 70\\
C6  & 22 & 0.109 & 0.007 & 0.100 & 0.123 & 52 & 0.138 & 0.006 & 0.128 & 0.154 & 74\\
C7  & 45 & 0.434 & 0.007 & 0.423 & 0.446 & 85 & 0.359 & 0.034 & 0.308 & 0.423 & 130\\
C8\textdagger  & 3  & -     & -     & -     & -     & 54 & 0.327 & 0.029 & 0.278 & 0.386 & 57\\
C9  & 26 & 0.084 & 0.009 & 0.074 & 0.108 & 52 & 0.095 & 0.013 & 0.073 & 0.119 & 78\\ 
C10 & 36 & 0.228 & 0.006 & 0.215 & 0.241 & 65 & 0.236 & 0.007 & 0.223 & 0.248 & 101\\
\hline
\rowcolor{white}\multicolumn{12}{l}{\textdagger No descriptive statistics due to lack or limited number of observations before team split.}\\
\end{tabular}
\end{table*}

\begin{figure*}[ht!]
\centering
  \includegraphics[width=.6\textwidth]{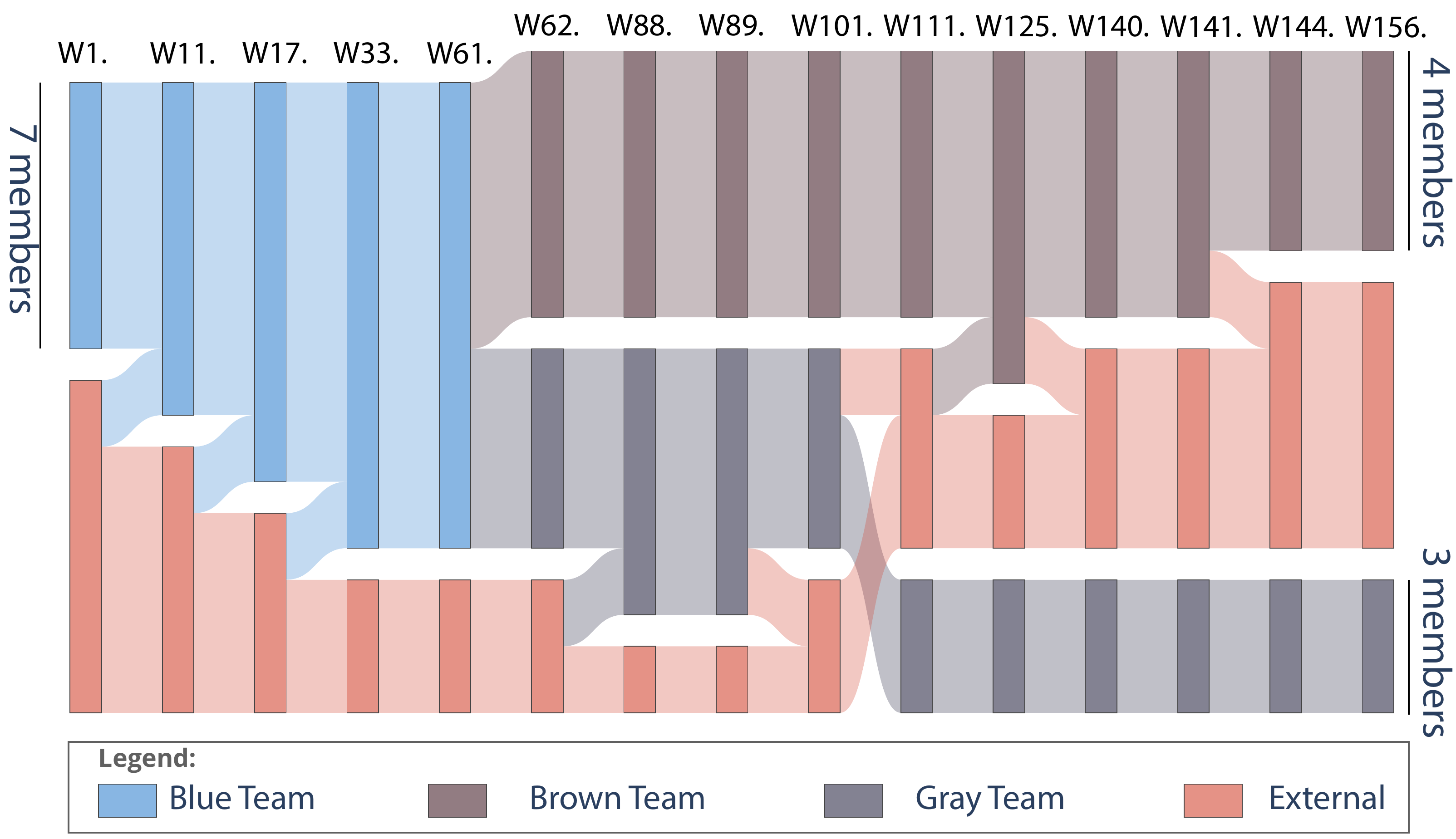}
\caption{Changes to the team during 2020 and 2022. The blue team split into two teams, team brown and team grey, on week 9 in 2021 (week 61).}
\label{fig:affiliationflow}      
\end{figure*}

As described in Section \ref{sec1:introduction}, the owning team of the components went through a drastic change and split into two teams on week 9 in 2021 (week 61). {\color{blue}Each new team took ownership of certain components. The decision was domain-driven, i.e., each team took the ownership of components within a certain domain.} Team Brown took ownership of components C1, C4, C6, C7, C9, and C10, while Team Gray took ownership of components C2, C3, C5, and C8 (see Table~\ref{tab:component_info}). There were other changes to teams' composition during the analysed period, i.e., members leaving a team or new members joining a team. The changes to the teams are illustrated in Figure~\ref{fig:affiliationflow}. 

\begin{figure*}[ht!]
\centering
  \includegraphics[width=.75\textwidth]{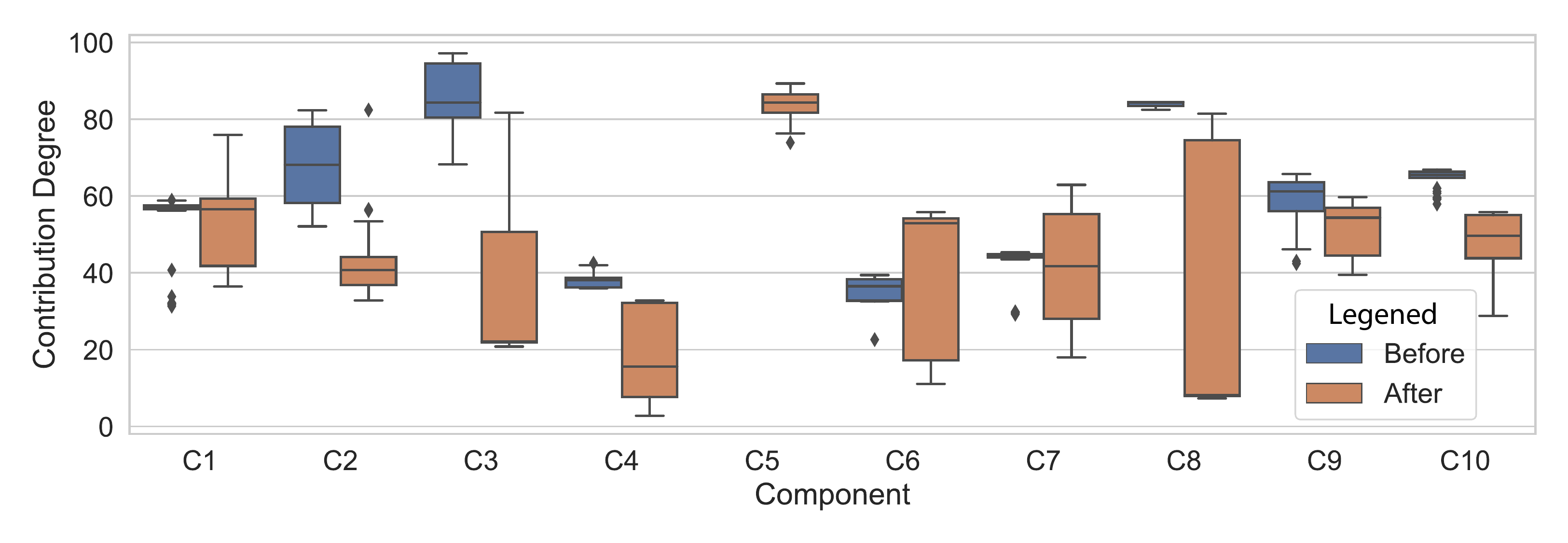}
\caption{Distribution of the \textit{Contribution Degree} observations for components `before' and `after' the split.}
\label{fig:box_contribution}      
\end{figure*}

\begin{figure*}[ht!]
\centering
  \includegraphics[width=.75\textwidth]{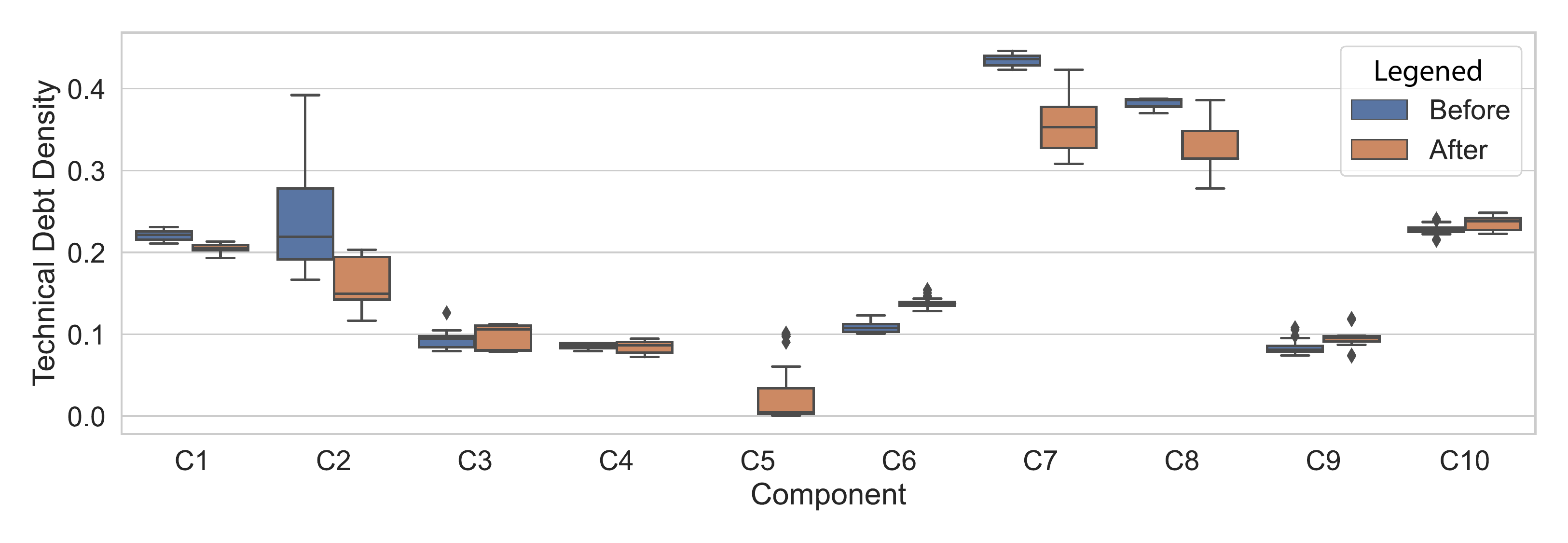}
\caption{Distribution of the TDD observations for components `before' and `after' the split.}
\label{fig:box_tdd}      
\end{figure*}

Figures \ref{fig:box_contribution} and \ref{fig:box_tdd} present the box plots for the distribution of \textit{Contribution Degree} and TDD respectively. 
We observe:

\begin{itemize}
    \item \noindent\textit{Contribution Degree}: The medians of distributions are different for components \textit{C2}, \textit{C3}, \textit{C4}, and \textit{C10}.
    \item \noindent\textit{TDD}: The medians of distributions are different for components \textit{C1}, \textit{C2}, \textit{C3}, \textit{C6}, \textit{C7}, \textit{C9} and \textit{C10}.
\end{itemize}

\begin{figure*}[ht!]
\centering
  \includegraphics[width=.82\textwidth]{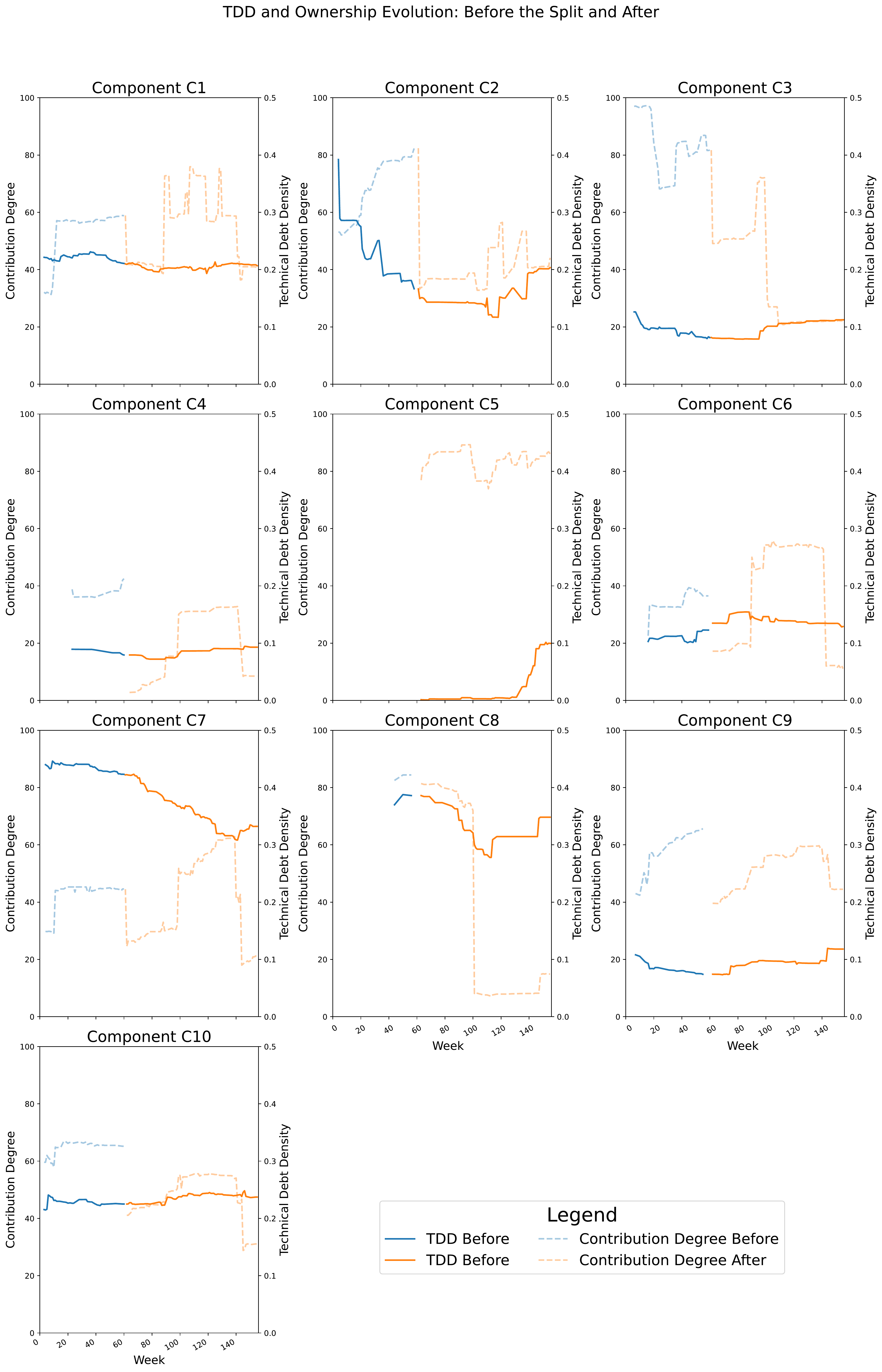}
\caption{The evolution of contribution degree and TDD in ten components.}
\label{fig:result_tdd_evo}      
\end{figure*}

Visual inspection of the components' changes to \textit{Contribution Degree} and TDD helps identify key events during the analysed period. These key events include the change in team structure, i.e., the team splitting into two teams on week 9 in 2021 (week 61) and key members leaving a team. For example, a {\color{blue}senior architect\footnote{Seniority is defined based on the number of years of experience in the field and in the company.} left Team Gray on week 37 in 2022 (week 141)}. Figure \ref{fig:result_tdd_evo} illustrates the evolution of \textit{Contribution Degree} and TDD for each component. The charts present the changes `before' and `after' the team split with blue and orange colours, respectively. The dashed lines illustrate the \textit{Contribution Degree}, and the solid line illustrates TDD.


\vspace{1mm}
\noindent\textbf{C1} [\texttt{Team Brown}]: The contribution degree for this component slightly increased before the team split. The split caused a sudden drop in contribution since some key developers for that component were allocated to Team Gray. Some weeks after that, the contribution degree went back to levels prior to the split and continued slightly growing. Conversely, TDD slightly decreased before and after the team split.

\vspace{1mm}\noindent\textbf{C2} [\texttt{Team Gray}]: We observe that TDD decreases as the contribution degree increases before and after the team split (the team keeping TDD under control). As in \textit{C1}, there is a major and sudden decrease in contribution degree after the team split. TDD slightly increases after the split.

\vspace{1mm}\noindent\textbf{C3} [\texttt{Team Gray}]: We observe that TDD slightly decreases as the contribution degree is high before the team split. As in most components, there is a major and sudden decrease in contribution degree when the team splits. And another major and sudden decrease in contribution degree was in week 101 when a product owner\footnote{This product owner was a senior developer for over 10 years working on the same components in the same team.} left the team. We observe an increasing trend of TDD during the period after the split. 

\vspace{1mm}\noindent\textbf{C4} [\texttt{Team Brown}]: We observe that the contribution degree for this component decreases significantly after the team split (close to $0\%$). TDD slightly decreases before the team split and increases significantly after the split. 

\vspace{1mm}\noindent\textbf{C5} [\texttt{Team Gray}]: This component was created after the team split. Therefore, we cannot compare it similar to the other cases. We can observe that the contribution degree is very high from the owning team ($80\%$ and higher) from the beginning, meaning that the owning team is its main contributor. As observed in Figure \ref{fig:result_tdd_evo}, TDD is very low at the beginning of the inception of the component. But it increases towards the end of the analysis, although the ownership levels remain in the same range.

\vspace{1mm}\noindent\textbf{C6} [\texttt{Team Brown}]: We observe that the contribution degree for this component decreases after the team split. And TDD starts to increase after the contribution degree drops. The owning team increases the contribution degree around week 80. This coincides with TDD slowly decreasing. There is a major and sudden decrease in contribution degree in week 141 when a senior architect left the team.

\vspace{1mm}\noindent\textbf{C7} [\texttt{Team Brown}]: This component has the highest amount of TDD among the investigated components. The component is a fork of an existing open-source repository that the team developed further internally. We observe that TDD decreases as the owning team increases its contribution degree. There is a major and sudden decrease in contribution degree in week 141 when a senior architect left the team.

\vspace{1mm}\noindent\textbf{C8} [\texttt{Team Gray}]: This component was created right before the team split. Contribution degree and TDD stay do not change significantly before and after the team split. There is a major and sudden decrease in contribution degree in week 101 when a product owner left the team. TDD started increasing after the product owner left the team.

\vspace{1mm}\noindent\textbf{C9} [\texttt{Team Brown}]: We observe that TDD decreases as the contribution degree increases before the team split. There is a sudden decrease in contribution degree after the team split. The owning team increased their contribution after the team split. TDD slightly increases after the split. 

\vspace{1mm}\noindent\textbf{C10} [\texttt{Team Brown}]: We observe that TDD has not changed substantially after the change. There is a decrease in contribution degree after the team split. However, the owning team has increased their contribution to the component after the split. 
There is a major and sudden decrease in contribution degree in week 141 when a senior architect left the team.


{\color{blue}
Before testing the potential effects of the contribution degree on TDD, we need to check whether the change in team structure (team split) affects the degree of ownership, i.e., whether we can observe statistically significant differences in the contribution degree before and after the split. If such differences are not deemed statistically significant, then we can study the impact of contribution degree on TDD without distinguishing before and after the split (since the factor split does not have an effect). However, if the differences are statistically significant, we will have to separate the two populations (before and after) since the factor team split does have an impact. 
}

Therefore, to test whether the change in the team structure significantly impacted contribution degree and TDD, we used the Mann-Whitney U test, as described in Section \ref{sec2:data_analysis}. The test results show that there is a significant difference between \textit{before} and \textit{after} the team split in contribution degree and TDD in most of the components (5 out of 8 for contribution degree and 7 out of 8 for TDD). Given that the team split is a confounding factor for analysing the association between contribution degree and TDD, we decided to separately perform Kendall's tau-b correlation test for the periods before and after the team split. The test results are presented in Table \ref{tab:mw_results}. The statistically significant cases are highlighted in bold.

\begin{table}[ht!]
\caption{Mann-Whitney U test results for contribution degree and TDD to compare the differences between \textit{before} and \textit{after} the teams split. Significant results are presented in boldface. The general .}
\label{tab:mw_results} 
\centering
\rowcolors{2}{white}{gray!25}
\begin{tabular}{lcc|cc}
 & \multicolumn{2}{c}{\textbf{Contribution Degree}} & \multicolumn{2}{c}{\textbf{Technical Debt Density}}\\
\rowcolor{white}& P-value & N  & P-value & N   \\
\hline
C1  & 0.571 & 122  & $<$\textbf{0.001} & 122  \\
C2  & $<$\textbf{0.001} & 89  & $<$\textbf{0.001} & 89  \\
C3  & $<$\textbf{0.001} & 105  & \textbf{0.029} & 105  \\
C4  & $<$\textbf{0.001} & 50  & 0585 & 50  \\
C5\textdagger & - & - & - & - \\
C6  & 0.195 & 74  & $<$\textbf{0.001} &  74 \\
C7  & 0.922 & 130  & $<$\textbf{0.001} & 130  \\
C8\textdagger & - & - & - & -  \\
C9  & $<$\textbf{0.001} & 78  & \textbf{0.001} & 78  \\
C10 & $<$\textbf{0.001} & 101  & $<$\textbf{0.001} & 101  \\
\hline
\rowcolor{white}\multicolumn{5}{l}{\textdagger No statistical test due to lack or limited number of observations}\\
\rowcolor{white}\multicolumn{5}{l}{before team split.}\\
\end{tabular}
\end{table}

The results from Kendall’s tau-b correlation coefficient for components 1-10 are presented in Table \ref{tab:test_results}. The results depict if there are any differences under two different occasions: before and after the team split. The statistically significant cases are highlighted in bold. The p-values less than 0.01 are identified with (\S), and p-values that are less than 0.05 are identified with (\textdagger).

The results for the correlation (see Table \ref{tab:test_results}), before the team split up, show that there is a statistically significant relationship between \textit{contribution degree} and TDD for components \textit{C1}, \textit{C2}, \textit{C4}, and \textit{C9}. For component \textit{C5}, we have no observations before the team split. We have very few observations for component \textit{C8}, i.e., 3 cases. Thus, we present no results for those two components before the team split.

On the other hand, in the results (see Table \ref{tab:test_results}) for the correlation after the team split, a statistically significant relationship between contribution degree and TDD was detected for all the components except for component \textit{C5}, where the team did not split, and component \textit{C9}.

\begin{table*}[ht!]
\caption{Test results - Kendall's $\tau$. Significant results are in boldface. Magnitude of association is calculated based on \cite{botsch2011significance}.}
\label{tab:test_results} 
\rowcolors{2}{white}{gray!25}
\centering
\begin{tabular}{lcccc|cccc}
 & \multicolumn{4}{c}{\textbf{Before}} & \multicolumn{4}{c}{\textbf{After}}\\
\rowcolor{white}& P-value & Kendall's $\tau$ & Magnitude & N & P-value & Kendall's $\tau$ & Magnitude & N  \\
\hline
C1 & \textbf{0.001} & \textbf{-0.332} \S & \textbf{Strong} & \textbf{46} & \textbf{0.013} & \textbf{-0.199} \textdagger & \textbf{Moderate} & \textbf{76} \\
C2 & $<$\textbf{0.001} & \textbf{-0.840} \S & \textbf{Strong} & \textbf{32} & \textbf{0.011} & \textbf{0.237} \textdagger & \textbf{Moderate} & \textbf{57} \\
C3 & 0.410 & 0.093 & Very Weak & 38 & $<$\textbf{0.001} & \textbf{-0.353} \S & \textbf{Strong} & \textbf{67} \\
C4 & \textbf{0.037} & \textbf{-0.529} \textdagger & \textbf{Strong} & \textbf{10} & \textbf{0.002} & \textbf{0.353} \S & \textbf{Strong} & \textbf{40} \\
C5* & - & - & - & - & 0.073 & 0.148 & Weak & 70 \\
C6 & 0.051 & -0.301 & Strong & 22 & \textbf{0.019} & \textbf{0.229} \textdagger & \textbf{Moderate} & \textbf{52} \\
C7 & 0.309 & -0.107 & Weak & 45 & $<$\textbf{0.001} & \textbf{-0.464} \S & \textbf{Strong} & \textbf{85} \\
C8* & - & - & - & 3* & $<$\textbf{0.001} & \textbf{0.770} \S & \textbf{Strong} & \textbf{54} \\
C9 & \textbf{0.001} & \textbf{-0.918} \S & \textbf{Strong} & \textbf{26} & 0.282 & 0.104 & Weak & 52 \\
C10 & 0.743 & 0.039 & Very Weak & 36 & $<$\textbf{0.001} & \textbf{0.454} \S & \textbf{Strong} & \textbf{65} \\
\hline
\rowcolor{white}\multicolumn{9}{l}{\textdagger Correlation is significant at the 0.05 level (2-tailed).}\\
\rowcolor{white}\multicolumn{9}{l}{\S Correlation is significant at the 0.01 level (2-tailed).}\\
\rowcolor{white}\multicolumn{9}{l}{* No statistical test due to lack or limited number of observations before team split.}\\
\end{tabular}
\end{table*}


\begin{figure*}[ht!]
\centering
  \includegraphics[width=.85\textwidth]{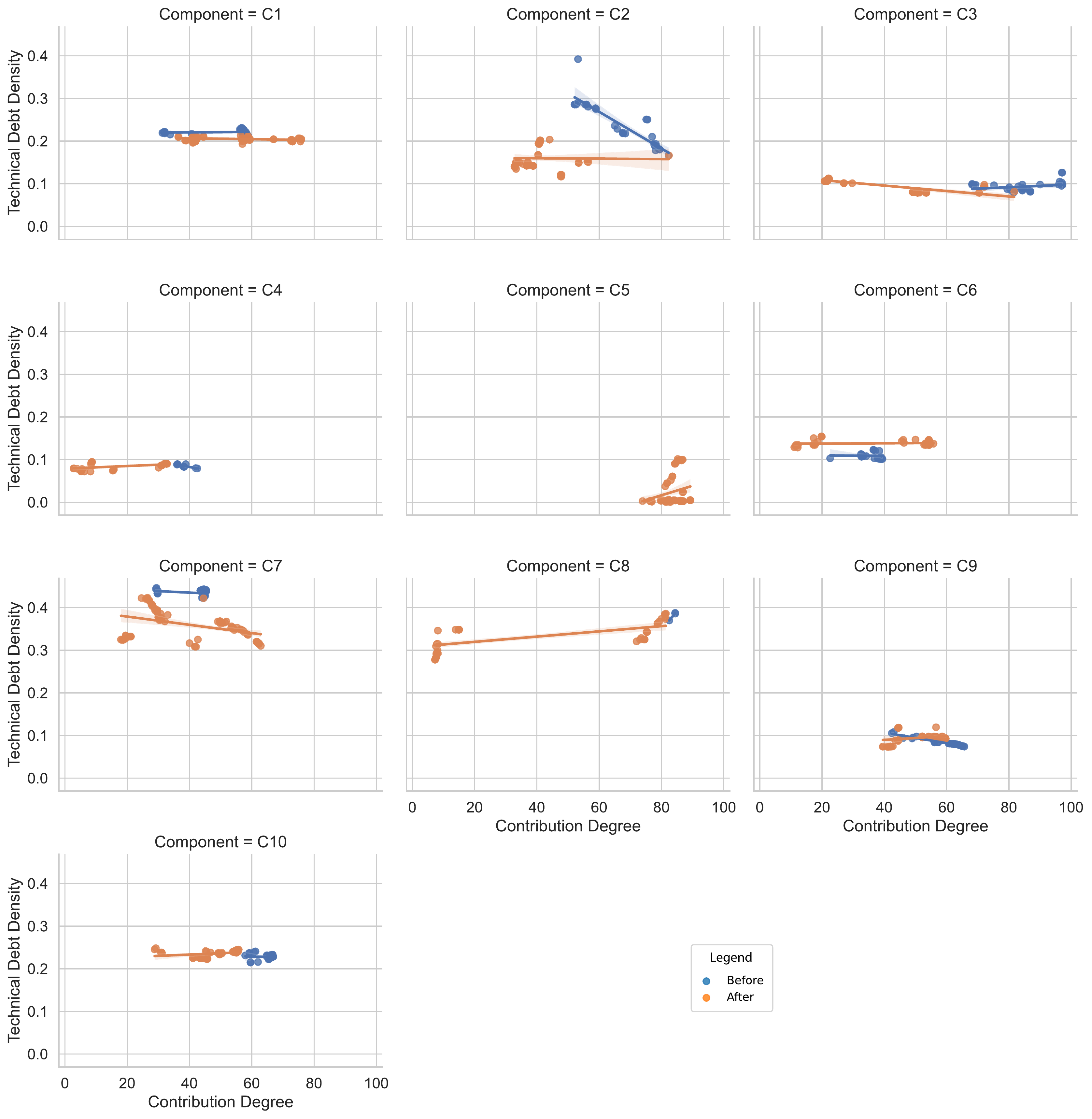}
\caption{Relationship between \textit{Contribution Degree} (X axis) and TDD (Y axis) for each component. The regression lines presented in the figure are only for visualisation. We are not using regression to describe the results nor prediction of contribution degree and TDD.}
\label{fig:result_contribution_degree}      
\end{figure*}

Figure \ref{fig:result_contribution_degree} illustrates the relationship between \textit{contribution degree} (X axis) and TDD (Y axis) for each component. 
We use Kendall’s tau-b correlation coefficient (see Table \ref{tab:test_results}) and Figure \ref{fig:result_contribution_degree} to investigate each component. Note that the regression lines presented in the figure are used only for visual inspection. We are not using regression to draw conclusions nor predict contribution degree and TDD.

\vspace{1mm}\noindent\textbf{C1} [\texttt{Team Brown}]: There is a significant strong negative correlation between contribution degree and TDD before the team split (P-Value: $0.001$; Kendall's $\tau$: $-0.332$) and a significant moderate negative correlation after team split (P-Value: $0.013$; Kendall's $\tau$: $-0.199$). Our observation and the test results suggest that as the contribution degree increases over time, TDD decreases in the component.
    
\vspace{1mm}\noindent\textbf{C2} [\texttt{Team Gray}]: There is a significant strong negative correlation between contribution degree and TDD before the team split  (P-Value: $<0.001$; Kendall's $\tau$: $-0.840$) and a significant moderate positive correlation after team split (P-Value: $0.011$; Kendall's $\tau$: $0.237$). We observe a different pattern after the team split. There is a significant moderate positive correlation between contribution degree and TDD; as contribution degree increases, TDD increases too. 
    
This increase in TDD might be due to the context of this particular component. The component is related to handling \textit{user authentication} in the system. There were issues integrating a third-party authentication application into the system and the increase in TDD can be attributed to the development of new features.

Furthermore, we observe that as the contribution degree decreases to $40\%$ and below, TDD changes unexpectedly, i.e., other factors such as responsibility diffusion \cite{Tornhill2015} and knowledge loss \cite{bosch2004software,tofan2011reducing} might have a bigger impact on TDD.

\vspace{1mm}\noindent\textbf{C3} [\texttt{Team Gray}]: There is no significant correlation between contribution degree and TDD before the team split (P-Value: $0.410$; Kendall's $\tau$: $0.093$). However, the results show a significant strong negative correlation after the team split (P-Value: $<0.001$; Kendall's $\tau$: $-0.353$). Our observation and the test results suggest a decrease in TDD as the contribution degree increases only before the split.

Similar to \textit{C2}, there is a decrease of contribution degree below $40\%$ around week 100 in this component, after which TDD starts increasing.

\vspace{1mm}\noindent\textbf{C4} [\texttt{Team Brown}]: There is a significant strong negative correlation between TDD and contribution degree before team split (P-Value: $0.037$; Kendall's $\tau$: $-0.529$) and a significant strong positive correlation after team split (P-Value: $0.002$; Kendall's $\tau$: $0.353$). Overall, the team's contribution degree is low in this component. 
    
We observe a different pattern after the team split with a significant strong positive correlation between contribution degree and TDD. This increase in TDD might be due to the context of this component, which is related to handling different \textit{licenses} in the system. There has been a significant amount of new development in this component, and the increase in TDD can be attributed to the development of new features while the owning team was gaining ownership.

We observe a decrease of contribution degree to near $0\%$ after the team split, after which TDD starts increasing.

\vspace{1mm}\noindent\textbf{C5} [\texttt{Team Gray}]: There is no significant correlation between TDD and contribution degree since the creation of this component (P-Value: $0.073$; Kendall's $\tau$: $0.148$).
    
\vspace{1mm}\noindent\textbf{C6} [\texttt{Team Brown}]: There is no significant correlation between TDD and contribution degree before the team split (P-Value: $0.051$; Kendall's $\tau$: $-0.301$). However, the results show a significant moderate positive correlation after the team split (P-Value: $0.019$; Kendall's $\tau$: $0.229$). There are two clusters for contribution degree observations ---orange dots--- after the team split (See Figure \ref{fig:result_contribution_degree} Component C6). This left cluster, i.e., lower contribution degree, is related to the fact that a senior architect left the team. The increase of the TDD after the team split is also associated with the same event.

\vspace{1mm}\noindent\textbf{C7} [\texttt{Team Brown}]: There is no significant correlation between TDD and contribution degree before the team split (P-Value: $0.309$; Kendall's $\tau$: $-0.107$). However, there is a significant strong negative correlation after the team split (P-Value: $<0.001$; Kendall's $\tau$: $-0.464$). Our observation and the test results suggest that as the degree of ownership increases over time after the split, TDD decreases in the component.

Similar to \textit{C2}, \textit{C3}, and \textit{C4}, we observe a decrease of contribution degree to below $40\%$ around week 145 in this component, after which TDD starts increasing.    

\vspace{1mm}\noindent\textbf{C8} [\texttt{Team Gray}]: There is a significant strong positive correlation between TDD and contribution degree after the team split (P-Value: $<0.001$; Kendall's $\tau$: $0.770$). There are two clusters for contribution degree observations ---orange dots---  after the team split (See Figure \ref{fig:result_contribution_degree} Component C8). The left cluster, i.e., lower contribution degree, is related to the removal of a product owner from the team. The increase of the TDD after the team split is also associated with the same event.

Similar to \textit{C2}, \textit{C3}, \textit{C4} and \textit{C7}, we observe a decrease of contribution degree to below $40\%$ around week 100 in this component, after which TDD starts increasing.

\vspace{1mm}\noindent\textbf{C9} [\texttt{Team Brown}]: There is a significant strong negative correlation between TDD and contribution degree before the team split (P-Value: $0.001$; Kendall's $\tau$: $-0.918$). However, there is no significant correlation after the team split (P-Value: $0.282$; Kendall's $\tau$: $0.104$). We observe a decrease of contribution degree to below $40\%$ after the team split, after which TDD starts increasing.
    
\vspace{1mm}\noindent\textbf{C10} [\texttt{Team Brown}]: There is no significant correlation between TDD and contribution degree before the team split (P-Value: $0.743$; Kendall's $\tau$: $0.039$). However, the results show a significant strong positive correlation after the team split (P-Value: $<0.001$; Kendall's $\tau$: $0.454$). There is a significant strong positive correlation between contribution degree and TDD. This increase in TDD is due to the context of this component. The component is related to handling \textit{users} in the system. The team maintains three versions of the same functionality at the same time, and the increase in TDD can be attributed to it.
\section{Discussion}\label{sec4:discussion}

This section discusses the insights gained from analysing the components and practical implications of the findings.


The findings of this study suggest that altering the structure of a team, specifically by dividing it into two separate teams, has the potential to impact the accumulation of technical debt. Specifically, our analysis reveals that components \textit{C2}, \textit{C4}, \textit{C6}, and \textit{C10} experienced a reduction in their contribution degree to below $40\%$, which corresponded with an increase in TDD in the subsequent weeks. Our observations are similar to the work presented by Nagappan et al. \cite{nagappan2008influence} on the impact of organisation structure on code quality.

The observations were further corroborated by the statistical tests conducted. However, it is worth noting that we cannot draw the same conclusion for component \textit{C8}, as its introduction occurred several weeks prior to the team split and thus was not subject to the same impact. These findings highlight the importance of carefully considering the potential effects of team restructuring on technical debt accumulation and the need for continued monitoring and analysis to ensure optimal team performance and productivity.

The results presented in Section \ref{sec3:results} suggest that contribution degree might impact the accumulation of TDD. We observe that before the split that in four components (\textit{C1}, \textit{C2}, \textit{C4}, \textit{C9})\footnote{For components \textit{C5} and C8 where there is no analysis before the split since \textit{C5} was a component created after the team split and \textit{C8} only a few weeks before, and therefore without enough data points to draw meaningful conclusions.} the contribution degree has a negative correlation, meaning that the higher the contribution degree, the lower TDD. After the split, we observe eight components for which the contribution degree seems to impact TDD. In components \textit{C1}, \textit{C3}, and \textit{C7}, we observe a statistically significant negative correlation between contribution degree and TDD; the higher the contribution degree, the lower TDD. However, we also observe that for components \textit{C2}, \textit{C4}, \textit{C6}, \textit{C8}, and \textit{C10}, there is a positive correlation, which might seem counter-intuitive. The first clarification is that a lower contribution degree does not mean not incurring in TD since many other factors might be impacting its accumulation, but what we observe, after inspecting Figures ~\ref{fig:result_tdd_evo} and \ref{fig:result_contribution_degree}, that TDD tend to grow together with contribution degree when the contribution degree is at levels below 40\% (except for the case of \textit{C5}), which might mean that controlling TD effectively is harder when the work in that component is mainly done by developers from outside the team.

One aspect to be considered when analysing the results after the split is that the change in trend on TDD might manifest with a few weeks of lag. In the first weeks after the split, the owning team might restrain themselves from making big changes, or the changes we observe might have been under development before the split and therefore have a different impact on TD. 

In the case of \textit{C5}, There might be several explanations, the first being the fact that during the first weeks of development for a component created from scratch, TDD is more easily kept under control, and when the functionality starts to grow TDD grows with it. We can also explain with the \textit{Technical Credit} concept. Experienced teams or individual developers might have \textit{credit} to incur in TD, similar to the financial concept of \textit{credit}, which allows a customer to loan money from a financial institution. Higher ownership might be linked with technical credit, which we will study in further work. C8 does not seem to follow the same pattern because \textit{C8} is created a few weeks before the split, but not from scratch. Instead, this component was created as a migration of a service in a monolith architecture to a microservices.\footnote{Most of the components under study are part of a migration from a monolith to a microservices architecture, but they count on a long development history. However, C8 is the only component that results from the migration of functionality from the monolith that occurred during the period under analysis.}

We observe that when a key contributor (senior architect and senior developer) leaves the team, the impact of ``architectural knowledge vaporisation'' \cite{tofan2011reducing,bosch2004software} might be manifested as an increase in technical debt. This result is aligned with the findings by Bird et al.~\cite{bird2011don}, which suggest that high levels of top contributors result in higher quality. We observed that when a senior developer left the team in Week 101, the contribution degree dropped more than $50\%$, in components \textit{C3} and \textit{C8}, resulting in an increase in TDD. Similarly, we observe the same when a senior architect left the team in Week 141, the contribution degree dropped more than $50\%$ in components \textit{C6} and \textit{C7}.

Our observations suggest that the events that cause major changes to contribution degree, i.e., the decrease of contribution degree, might impact the accumulation of TDD. As reported by \cite{de2018measuring}, software quality is directly linked to the volume of collaboration and commitment of the development team. Therefore, the impact of a major decrease in contribution degree can be manifested as the faster accumulation of technical debt (i.e., TDD growth over time).


\subsection*{\textbf{Implications for Research and Practice}}

\begin{itemize}
    {\color{blue}
    \item When assigning component ownership to teams, it is crucial to consider the contribution degree and its potential impact on the accumulation of TD to each component. The changes in contribution degree can arise from a number of factors, including organisational changes ---changes to team constellation--- and attrition ---team members leaving a team or the company---, among others. Neglecting these factors when making component ownership decisions can lead to faster accumulation of TD, which impacts development effectiveness and efficiency. Therefore, managers and team leaders must carefully evaluate the degree of contribution to each component.
    
    \item In the context of organisational restructuring, when the creation of new teams involves splitting existing ones, a key consideration is assigning, when possible, components to the sub-team that exhibits the highest degree of contribution. This rationale may be based on a number of factors, such as the distribution of expertise and skills among team members, the complexity of the component, or the urgency of the task at hand. By assigning components in this manner, organisations can aim to minimise the impact of the uncontrolled accumulation of TD.


    \item The contributions of the senior members of the team are important to preserving the quality of the code \cite{posnett2013dual}. Attrition is unavoidable and members of the team might leave due to a variety of reasons. Companies and development teams should understand the degree of contribution of individual team members to try to preserve knowledge and avoid architectural knowledge vaporisation before key players leave the organisation or the teams. 
    }
    {\color{blue}
    \item In general, high contribution degree levels seem to be a good way to manage TD effectively. TD may accumulate over time due to responsibility diffusion. The \textit{responsibility diffusion} refers to the decreased responsibility each individual (i.e., team member) feels when they are part of a (large) group of people contributing to the same piece of code. If there are several teams contributing to a repository, and the owner has a low degree of contribution, owners might either experience the responsibility diffusion effect when observing TD items (somebody else will fix them) or not have enough knowledge to fix them. Thus, we argue that a higher degree of contribution in the owning team (i.e., the main contributor team or a team that contributes more than other teams to a particular component) could be a way to tackle responsibility diffusion. In other words, multiple teams might be contributing to a particular component. However, if the main contributing team is also the owning team, they will have the knowledge to fix TD items, and they might (consciously) put more effort into caring about the code of a particular component and perform TD mitigation activities, e.g., bug fixing or refactoring~\cite{sundelin2021towards}.
    }

    {\color{blue}
    \item Ownership and contribution alignment and its implications (i.e., TD) are phenomena that need further study, especially in proprietary software development. This might also help us further understand the potential impact of organisational changes. 

    }
\end{itemize}

\section{Limitations and Threats to Validity}\label{sec5:threats_to_validity}

Different types of threats to validity might impact the results of our research. \textit{Construct validity} concerns the selection of measurements that reflect the constructs. The study constructs and measurements used in this study are presented in Section~\ref{sec2:research_methodology}. We use several measurements from different tools including \texttt{Git}, \texttt{BitBucket}, \texttt{Jira}, and \texttt{SonarQube}. 

{\color{blue}Using research tools can threaten construct validity due to measurement errors, operationalisation issues, instrument bias, respondent interpretation, and external influences. However, to mitigate these threats, we have considered the following: involving expert reviews from the team when reviewing the data, ensuring reliability (details below), using clear operational definitions as decided by the company, and standardising administration (company-wide use of the tool). These steps help ensure that the tool accurately measures the intended construct within the context and that the results are consistent, leading to more valid and reliable research outcomes. 

We are aware of the limitations of static analysis tools (e.g., SonarQube) in detecting TD items \cite{Lenarduzzi2023}, however first, SonarQube is the tool (and construct) that the studied company (and many others) uses for reasoning about technical debt and code and architectural degradation. Second, it has been used in other studies approaching the evolution of Technical Debt \cite{siavvas2022technical,digkas2020can,lenarduzzi2020sonarqube,zabardast2020,lenarduzzi2019diffuseness,digkas2018developers,digkas2017evolution}. 
We use \textit{TDD}~\cite{al2019evolution} extracted from SonarQube and \textit{contribution degree} calculated using OCAM~\cite{zabardast2022}. The former has been used in other studies to investigate the impact of external factors such as Clean Code \cite{digkas2018developers}, and the latter is an extension of the model presented in \cite{zabardast2022}. 

We do not distinguish between the TD items when conducting the analysis, i.e., all TD items are considered to have the same weight in the calculations. The repaid TD is calculated through the issues flagged as `fixed' and `closed'. However, there might be cases where TD has been removed by deleting the files that might not be included in our analysis. 

Threats to construct validity in using the OCAM model for measuring the degree of alignment between ownership and contribution may include incorrect or imprecise definitions of key concepts, measurement errors, and biases in data collection. To mitigate these threats, it is essential to establish clear, consistent definitions for ownership and contribution, use reliable and objective measurement tools, and validate these tools through pilot testing and feedback. Ensuring diverse stakeholder involvement in defining and refining the constructs can also help improve validity.
}

We mitigated the potential threats regarding the calculations of measures and the selection of cases in consensus with the heads of development in the studied company. 

\textit{Internal validity} concerns the extent to which the results are free from error, i.e., the results represent the truth. A major threat to the validity of this study is the presence of confounding factors that, together with ownership and contribution misalignment, might impact the results presented in this work. We present the results from correlation tests, i.e., the results presented in this article do not aim to provide a cause-and-effect relationship between the studied variables. We mitigate this threat to validity by providing contextual information regarding the components to find other possible factors impacting our observations. There might be other non-studied factors that can explain the observations. Additional studies are required to better understand the other factors impacting the growth of TDD. 

\textit{External validity} concerns the generalisability of the results. The study results presented in this article only apply to the components investigated in this case study. The case study results should not be considered generalisable based on cases. We use statistical tests strictly to reason about each specific case.

\textit{Reliability}, as one of the biggest threats to validity in this study, concerns the collected data and the performed analysis to be independent from individual researchers. We mitigated this threat to validity by consistently and frequently involving the developing teams ---holding bi-weekly meetings--- throughout the study design and interpreting the results.

Finally, \textit{conclusion validity} refers to the degree to which the conclusions drawn from a statistical analysis are accurate and reliable \cite{Sjoberg2022}. It is concerned with whether the statistical inferences are based on a sound and robust data analysis. To mitigate this threat, we used appropriate statistical methods and considered potential confounding variables that could affect the results, i.e., team split. We also collected an adequate sample size representative of the population being studied.

\section{Conclusion}\label{sec7:conclusion}

In this article, we present the results of an industrial case study on the impact of ownership and contribution alignment on code technical debt. We investigated ten components during a three-year period (2020 to 2022) from a large software development company that develops web-based financial and accountancy services.

Study results reveal that prior to the team structure's change, a negative correlation was found between contribution degree and TDD in the majority of cases where high levels of contribution degree were present. This negative correlation means that higher contribution degree correlate with lower TDD was statistically significant in four components, indicating that heightened contribution degree was associated with decreased TDD. Once the original team was divided into two, a statistically significant negative correlation was observed in three components, whereas five components exhibited a positive correlation that may be attributed to low levels of contribution degree, thereby suggesting the team's inability to effectively manage TDD.

The results suggest that, when assigning the components' ownership responsibility to teams, it is important to consider the contribution degree of each team to make sure that the team receiving this responsibility is ready to take the required actions to preserve the components' TDD under control. Additionally, the contributions of senior team members might be crucial to mitigating the accumulation of code TD and preserving knowledge when attrition occurs. High levels of contribution degree can help manage TD effectively by restraining responsibility diffusion and encouraging the owning team to invest more effort in mitigating TD. The study highlights the need for further research on the ownership and contribution alignment and their impact on TD in proprietary software development, which may help organisations better understand the potential effects of organisational changes.

However, both concepts, ownership and contribution degree alignment and TD, are complex phenomena, and other factors might impact the changes in the accumulation of TD. In addition, we are aware that the results are not generalisable, but the aim of this paper is also to raise awareness of the importance of aligning ownership and contribution degree. 

We believe the results presented in this article need to be strengthened with future replications of the study. Finally, we plan to investigate the concept of TD credit in other cases.

\section*{Acknowledgements}
This research was supported by KK foundation through the KKS Profile project SERT 2018/010 at Blekinge Institute of Technology, Sweden.

\bibliographystyle{IEEEtran}
\bibliography{references}

\end{document}